\documentstyle[prc,aps,epsfig]{revtex}
\begin{document}
\draft
\title{Effect of the relativistic spin rotation
       on two--particle spin composition}
\author{R. Lednicky$^{1,2}$, V.L.Lyuboshitz$^{1}$
        and V.V.Lyuboshitz$^{1}$}
\maketitle
\maketitle
\begin{center}
{\small {\it
$^{1}$ JINR, 141980 Dubna, Moscow region, Russia.
 \\[0pt]
$^{2}$ Institute of Physics ASCR, Na Slovance 2, 18221 Prague 8, Czech
Republic.\\[0pt]
}}
\end{center}

\begin{abstract}
The effect of the relativistic spin rotation
on two--particle spin states,
conditioned by the
setting of the spins of the particles in their rest frames and by the
noncommutativity of the Lorentz transformations along noncolinear
directions, is discussed.
Particularly, the transition from the c.m.s. of two spin-1/2 particles
to the laboratory is considered.
When the vectors of the c.m.s. particle velocities are not
colinear with
the velocity vector of the c.m.s., the angles of the
relativistic spin rotation for the two particles
are different. As a result, the relative fractions
of the singlet and triplet states in the relativistic system of two
spin-1/2 particles with a nonzero vector of relative momentum
depend on the concrete frame in which
the two-particle system is analyzed.
\end{abstract}
\vspace{0.5cm}

{\bf 1}. Earlier the spin correlations in two-particle quantum systems were
analyzed in detail as a tool allowing one to measure the space--time
characteristics of particle production [1-5], to study the two--particle
interaction and the production dynamics (see \cite{led99,ll01} and
references therein)
and to verify the consequences
of the quantum--mechanical coherence with the help of Bell--type
inequalities \cite{ll01}.

The spin state of the system of two particles
in an arbitrary frame
is described by the two-particle density matrix, the elements of
which,
$
       \rho^{(1,2)}_{{m_1 m_1}';\,{m_2 m_2}'},
$
are given in the representation of the spin projections of the first
and second particle in the corresponding rest frames onto the
common coordinate axis $z$
(see, {\it e.g.}, \cite{led99,lll03}).\footnote
{The setting of the particle spins in their respective rest frames
is based on the properties of the inhomogeneous Lorentz group and
avoids the problem of the noncommutativity of the spin operators
with the free Hamiltonian \cite{fol56}. This circumstance was not
understood in Ref. \cite{al95},
where the unnecessary condition of nonrelativistic particle velocities
was required.}
However, one should take into account the relativistic spin rotation
conditioned by the
additional rotation of the spatial axes at the successive
Lorentz transformations along
noncolinear directions [7-9].\footnote{
The relativistic
rotation of the spatial axes leads to the nontransitivity
of the parallelism in the theory of relativity (see \cite{shi99}
and references therein): generally, the parallel axes of the
frames $K_1$ and $K$, $K_2$ and $K$ do not imply the parallel axes
of the frames $K_1$ and $K_2$. The axes of all the three frames
could be mutually parallel if only their velocities were colinear
(for example, if $K_1$ and $K_2$ were the rest frames of the two
particles and $K$ - their c.m.s.).
}
As a result,
the {\it concrete} description of a particle spin state
depends on the frame from which the transition to the
particle rest frame is performed.
Particularly, the total spin composition of the two--particle
state with a nonzero vector of relative momentum
is generally frame--dependent due to different
relativistic rotation angles of the two spins at the transition
to the frame moving in the direction which is not colinear with
the velocity vectors of both particles.

Usually, it is convenient to consider the spin correlations in the
center-of-mass system (c.m.s.) of the particle pair.
This is natural at the addition of the two--particle total spin
and the relative orbital angular momentum into the
conserved total angular momentum.
In some cases, however, it may be useful to make transition to
the laboratory, {\it e.g.}, in the case when the particle
scatterings are used as their spin analyzers \cite{lp97}.\footnote
{In principle, this transition is not necessary since one can
transform the four--vectors defining the
polarization analyzers first to the pair c.m.s.
and then to the respective particle rest frames.}
Denoting $M_l$ and
${\bf p}_l = \pm {\bf k}$ the masses and
c.m.s. momenta of the two particles, $l=1,2$,
their respective c.m.s. velocities in the units of the velocity
of light ($c=1$) are
${\bf v}_l = \pm {\bf k}/ \sqrt{{\bf k}^2 +
M_l^2}$. Here and below the $\pm$ signs correspond to the
first ($l=1$) and second ($l=2$) particle, respectively.
We denote the corresponding laboratory velocities as
$\widetilde{{\bf v}}_l$
and the laboratory velocity of the particle pair as ${\bf V}$.
At the Lorentz transformation
from the c.m.s. of the particle pair to the laboratory frame
with parallel respective spatial axes, the spins
of the first and the second particle (in their respective rest frames)
rotate in opposite directions around the axis which is parallel
to the vector ${\bf [kV]}$.\footnote{
The relativistic spin rotation is
the purely kinematical effect:
the angles of the space rotation coincide with the angles
between the vectors of the resulting velocities at the relativistic
addition of velocities ${\bf v}_{l}$ and ${\bf V}$ in the direct
and reverse orders \cite{bgmr68}.
}
The rotation angles $\omega_l$ are given by the Stapp formula
\cite{sta56} (see also \cite{cs58,rit61}):
\begin{equation}
 \sin \omega_l = \pm \gamma \gamma_l V v_l  \sin \theta
\frac { 1 + \gamma + \gamma_l + \widetilde{\gamma}_l}
{(1 + \gamma) (1 + \gamma_l) ( 1+ \widetilde{\gamma}_l)},
\end{equation}
where the positive sign corresponds to the direction of the
nearest rotation from the vector ${\bf k}$ to the vector ${\bf V}$;
$\theta$ is the angle between the vectors ${\bf k}$ and ${\bf V}$
\,($0 \le \theta \le \pi$),
$v_l = |{\bf v}_l|$, $\widetilde{v}_l= |\widetilde{{\bf v}}_l|$,
$V =|{\bf V}|$ and
  $\gamma_l = \left(1 - v_l^2\right)^{-1/2}$,
$\widetilde{\gamma}_l = \left(1 - {\widetilde v}_l^2\right)^{-1/2}$,
$\gamma = \left( 1 - V^2 \right)^{-1/2}$
are the Lorentz factors;
\begin{equation}
 \widetilde{\gamma}_l
= \gamma_l \gamma \left(1 \pm v_l V \cos \theta\right).
\end{equation}
In the case of equal--mass particles the relations ${\bf v}_1 =
-{\bf v}_2,\, \gamma_1 = \gamma_2$ hold (but
 $\widetilde{\gamma}_1 \ne\widetilde{\gamma}_2$ when $V\cos\theta\ne 0$).

Using the equality
$$
  (1 + \gamma + \gamma_l + \widetilde{\gamma}_l)^2 =
2(1 + \gamma) (1 + \gamma_l) (1 +  \widetilde{\gamma}_l)
- (\gamma^2 - 1) (\gamma_l^2 - 1)  \sin^2 \theta,
$$
one can write the analogous expressions for
the cosines of the spin rotation angles:
$$
\cos\omega_l = 1 - \frac{(\gamma - 1)(\gamma_l  - 1)}
{(1 +  \widetilde{\gamma}_l)}\,\sin^2\theta.  \eqno(1a)
$$
In the case of the colinearity of the velocity vectors ${\bf v}_l$
and ${\bf V}$, when $\theta =0$ or $\theta = \pi$,
both the rotation angles are equal to zero.

At nonrelativistic velocities $v_l$ in the c.m.s. of the particle pair
($\gamma_l\approx 1$, $\widetilde{\gamma}_l\approx\gamma$), the
angles $\omega_{l}$ of the spin rotation are small and scale with
$v_l$:
\begin{equation}
\omega_l \approx \pm
\frac{\gamma}{\gamma + 1}v_l V \sin\theta.
\end{equation}
In the ultrarelativistic limit, when $\gamma_l \rightarrow \infty$,
$\widetilde{\gamma}_l/\gamma_l \rightarrow \gamma \left(
 1 \pm V \cos\theta\right)$, one has
\begin{equation}
\sin \omega_l \approx \pm V \sin \theta \frac{1 + \gamma
(1 \pm V\cos \theta)}{(1 +\gamma)(1 \pm V\cos \theta)},\quad
\cos\omega_l \approx 1  - \frac{\gamma -1}{\gamma (1
 \pm V\cos\theta)}\sin^2\theta.
  \end{equation}
The relations (4) are valid exactly for massless
particles (photons, neutrinos). In this case
the rotation angles coincide with the aberration angles
(the angles between the vectors ${\bf v}_l$ and
$\widetilde{{\bf v}}_l$); then
the helicity (the spin projection of the particle onto the direction
 of its momentum) is the relativistic invariant \cite{rit61}.

Taking into account the relativistic spin rotation
at the transition from the two--particle c.m.s.
to the laboratory, the two--particle spin
density matrix is transformed as follows:
\begin{equation}
\hat{\rho}'^{(1,2)} = \hat{D}^{(1)}(\omega_1) \otimes \hat{D}^{(2)}(\omega_2)
\hat{\rho}^{(1,2)} \hat{D}^{(1)+}(\omega_1) \otimes \hat{D}^{(2)+}(\omega_2),
\end{equation}
where
\begin{equation}
\hat{D}^{(l)}(\omega_l)= \exp (i\omega_l {\bf \hat{j}}_l{\bf n})
\end{equation}
are the matrices of the space rotations generated by the
vector spin operators $\hat{{\bf j}}_l$, ${\bf n}$ is the unit vector
parallel to the direction of the vector $[{\bf kV}]$.

{\bf 2}. In the case of two spin--1/2 particles, the two-particle
spin density matrix has the structure \cite{lp97,led99,ll01}:
\begin{equation}
\label{eq7}
\hat{\rho}^{(1,2)} =\frac{1}{4} [\hat{I}^{(1)} \otimes
\hat{I}^{(2)} + (\hat{\mbox{\boldmath $\sigma$}}^{(1)} {\bf P}_1)
\otimes \hat{I}^{(2)} + \hat{I}^{(1)} \otimes
(\hat{\mbox{\boldmath $\sigma$}}^{(2)} {\bf P}_2) + \sum_{i=1}^3
\sum_{k=1}^3 T_{ik}\hat{\sigma}_i^{(1)} \otimes
\hat{\sigma}_k^{(2)}].
\end{equation}
Here $ \hat{I}$ is the two-row unit matrix,
$\hat{\mbox{\boldmath $\sigma$}}$ is the Pauli vector operator,
${\bf P}_l= \langle \hat{\mbox{\boldmath $\sigma$}}^{(l)}\rangle$
are the polarization vectors,
$ T_{ik}= \langle
\hat{\sigma}_i^{(1)} \otimes \hat{\sigma}_k^{(2)}\rangle $
are the components of the correlation tensor, $ \{1,2,3\} \equiv \{x,y,z\}$.
The left and right indexes of the correlation tensor
correspond to the rest frames of the first ($l=1$) and second ($l=2$)
particle, respectively.
The corresponding probability to select the particles
with the polarizations $\mbox{\boldmath $\zeta$}^{(l)}$
can be obtained by
the substitution of the matrices $\hat{\sigma}_i^{(l)}$
in the expression (\ref{eq7}) with the corresponding
projections $\zeta_i^{(l)}$.
Particularly, when analyzing the polarization states with the
help of particle decays, the vector analyzing power
$\mbox{\boldmath $\zeta$}^{(l)}=\alpha_l {\bf n}_l$,
where $\alpha_l$ is the decay asymmetry corresponding to
the decay analyzer unit vector ${\bf n}_l$.
As a result \cite{led99,ll01}, the correlation between the decay analyzers
is determined by the product of the decay asymmetries and the
trace of the spin correlation tensor
$$T=T_{xx}+T_{yy}+T_{zz}.$$
For example, the angular correlation ${\bf n}_1{\bf n}_2=\cos\theta_{12}$
between the directions of the three--momenta of the decay protons in the respective
rest frames of two $\Lambda$-hyperons decaying into the channel
$\Lambda \rightarrow p + \pi^-$ with the $P$-odd asymmetry
$\alpha = 0.642$ is described by the normalized probability density
 \begin{equation}
 \label{eq7a}
 W(\cos\theta_{12}) = \frac{1}{2} \left( 1 + \alpha^2\frac{T}{3}
 \cos\theta_{12} \right).
 \end{equation}

Clearly, the structure of both Eq. (\ref{eq7}) and the corresponding
angular distribution of the spin analyzers (e.g., Eq. (\ref{eq7a}))
does not depend on the system from which the transitions to the
particle rest frames are performed. The system dependence manifests
only through the relativistic rotations in the successive Lorentz
transformations along noncolinear directions.
The matrices of the space rotations due to the transition from
the c.m.s. of two free spin-1/2 particles to the laboratory are
the following:
\begin{equation}
\label{eq8}
\hat{D}^{(l)}(\omega_l) = \cos \frac{\omega_l}{2} +
 i \hat{\mbox{\boldmath $\sigma$}}^{(l)} {\bf n}
\,\sin \frac{\omega_l}{2}.
\end{equation}
Selecting the $z$--axis parallel to the direction of the vector
${\bf n}=[{\bf kV}]/|[{\bf kV}]|$,
and the axes $x$ and $y$ in the plane perpendicular to this
vector, the polarization vectors and the spin correlation tensor
transform at the transition to the laboratory
in accordance with the (active) rotations around the
z--axis by the angles $\omega_1$ and $\omega_2$
for the first and second particle, respectively:

\begin{equation}
\label{eq9}
\begin{array}{c}
P'_{lx}=P_{lx}\cos\omega_l - P_{ly}\sin\omega_l;\quad
  P'_{ly}=P_{ly}\cos\omega_l + P_{lx}\sin\omega_l;\quad
  P'_{lz}=P_{lz};
\\ \\
T'_{xx} = (T_{xx}\cos\omega_1 - T_{yx}\sin\omega_1)\cos\omega_2 -
            (T_{xy}\cos\omega_1 - T_{yy}\sin\omega_1)\sin\omega_2;
\\ \\
T'_{yy} = (T_{yy}\cos\omega_1 + T_{xy}\sin\omega_1)\cos\omega_2 +
            (T_{yx}\cos\omega_1 + T_{xx}\sin\omega_1)\sin\omega_2;
 \quad T'_{zz} = T_{zz};
 \\ \\
T'_{xy} = (T_{xy}\cos\omega_1 - T_{yy}\sin\omega_1)\cos\omega_2 +
            (T_{xx}\cos\omega_1 - T_{yx}\sin\omega_1)\sin\omega_2;
\\ \\
T'_{yx} = (T_{yx}\cos\omega_1 + T_{xx}\sin\omega_1)\cos\omega_2 -
            (T_{yy}\cos\omega_1 + T_{xy}\sin\omega_1)\sin\omega_2;
\\ \\
T'_{xz} = T_{xz} \cos\omega_1 - T_{yz} \sin\omega_1;
 \qquad  T'_{zx} = T_{zx} \cos\omega_2 - T_{zy} \sin\omega_2;
\\ \\
T'_{yz} = T_{yz} \cos\omega_1 + T_{xz} \sin\omega_1; \qquad
T'_{zy} = T_{zy} \cos\omega_2 + T_{zx} \sin\omega_2.
\end{array}
\end{equation}
Particularly, the trace of the spin correlation tensor transforms
at the transition to the laboratory as:
\begin{equation}
\label{eq10}
T' = (T_{xx} + T_{yy}) \cos(\omega_1-
\omega_2) + (T_{xy} - T_{yx}) \sin(\omega_1 - \omega_2) + T_{zz}
\end{equation}
or,
in the case of a symmetric tensor, as:
\begin{equation}
\label{eq11}
T'= T - 2\,(T_{xx} + T_{yy})\,\sin^2 \frac{\omega_1 -\omega_2}{2}.
\end{equation}
So, the c.m.s. trace $T$ in Eq. (\ref{eq7a}) is substituted by the
laboratory one $T'$ calculated using Eq. (\ref{eq10}) or
(\ref{eq11}) together with Eqs. (1) and (1a) for the spin
rotation angles.

{\bf 3}. It was shown \cite{led99,ll01} (see also
\cite{jay78,bar91})
that the trace of the
correlation tensor of a system of two spin-1/2 particles is the
following linear combination of the relative fractions of singlet
(the total spin $S=0$) and triplet ($S=1$) states:
\begin{equation}
\label{eq13}
 T=\langle \hat{\mbox{\boldmath $\sigma$}}^{(1)} \otimes
 \hat{\mbox{\boldmath $\sigma$}}^{(2)}\rangle =
\rho_t -3\rho_s, \qquad \rho_t + \rho_s =1.
\end{equation}
When we have the pure singlet state of the particle pair in
its c.m.s. ($\rho_s=1,\,\rho_t=0, T_{ik} = - \delta_{ik},\, T=-3$),
the transformation to the laboratory gives
\begin{equation}
\label{eq14}
   T' =  -3 + 4 \sin^2 \frac{\omega_1 - \omega_2}{2}.
\end{equation}
It follows from Eqs. (\ref{eq13}) and (\ref{eq14}) that at the
transition to the laboratory the relative fraction
of the singlet state decreases in favor of a triplet state:
\begin{equation}
\label{eq15}
 \rho'_s = \cos^2 \frac{\omega_1 - \omega_2}{2}, \qquad
 \rho'_t = \sin^2 \frac{\omega_1 - \omega_2}{2}.
\end{equation}
Thus the square of the total spin of two free particles
with a nonzero vector of relative velocity is
not a relativistic invariant (see \cite{cs58}).
Introducing the two--particle singlet state:
\begin{equation}
|\psi\rangle_{00} =
 \frac{1}{\sqrt{2}} \left(|+1/2\rangle^{(1)}_z \, |-1/2\rangle^{(2)}_z
- |-1/2\rangle^{(1)}_z \,
|+1/2\rangle^{(2)}_z \right)
\end{equation}
and the triplet state with the zero projection onto the rotation
axis $z$:
\begin{equation}
|\psi\rangle_{10} =
 \frac{1}{\sqrt{2}} \left(|+1/2\rangle^{(1)}_z \, |-1/2\rangle^{(2)}_z
+ |-1/2\rangle^{(1)}_z \,
|+1/2\rangle^{(2)}_z \right),
\end{equation}
the result in Eq. (\ref{eq15}) also follows directly
from the matrices of space rotations in Eq. (\ref{eq8});
the singlet state in the two--particle c.m.s.
is transformed into the following superposition of the singlet
and triplet states in the laboratory:
\begin{equation}
\label{eq17}
|\psi'_s\rangle =
\cos\frac{\omega_1 - \omega_2}{2}|\psi\rangle_{00}
+ i \sin\frac{\omega_1 - \omega_2}{2}|\psi\rangle_{10}.
\end{equation}

Similarly, the transformation of the pure
triplet state $|\psi\rangle_{10}$ in the two--particle
c.m.s.
 ($\rho_s=0,\, \rho_t=1,\,  T_{zz}=  -1, T_{xx} =T_{yy}=1, T=1$)
to the laboratory gives
  \begin{equation}
  T'= 1 - 4 \sin^2\frac{\omega_1 - \omega_2}{2},
  \end{equation}
the corresponding fractions being
  \begin{equation}
  \rho'_s = \sin^2 \frac{\omega_1 - \omega_2}{2}, \qquad
 \rho'_t = \cos^2 \frac{\omega_1 - \omega_2}{2},
  \end{equation}
in accordance with the transformation:
\begin{equation}
\mid\psi'_t\rangle = \cos\frac{\omega_1 -
\omega_2}{2}|\psi\rangle_{10} + i \sin\frac{\omega_1 -
\omega_2}{2}|\psi\rangle_{00}.
\end{equation}

In the case of the unpolarized triplet in the two--particle c.m.s.
($\rho_s=0,\, \rho_t = 1,\, T_{ik} =
\delta_{ik}/3,\, T=1$ \cite{lp97,ll01}), we have
 \begin{equation}
T' = 1 - \frac{4}{3} \sin^2 \frac{\omega_1 - \omega_2}{2},
\end{equation}
\begin{equation}
 \rho'_s = \frac{1}{3} \sin^2\frac{\omega_1 - \omega_2}{2},\qquad
 \rho'_t = 1 - \frac{1}{3} \sin^2\frac{\omega_1 - \omega_2}{2}.
\end{equation}

Using Eqs. (1) and (1a), it is easy to show that in the case of
two spin-1/2 particles with the same masses ($\gamma_2 = \gamma_1,\,
v_2 = v_1$) the measure of the spin mixing
can be written in the form:
\begin{equation}
   \kappa \equiv \sin^2 \frac {\omega_1 - \omega_2}{2}
   = \left (v_1 V\right )^2 \sin^2\theta
\left [\left (\frac{1}{\gamma}
 + \frac{1}{\gamma_1}\right)^2 + \left( v_1 V\right)^2
 \sin^2 \theta \right ]^{-1}.
\end{equation}
The maximum of the mixing factor $\kappa$ corresponds to the angle
$\theta = \pi/2$. In the ultrarelativistic limit, when
$\gamma_1\gg 1,\, \gamma\gg 1$ and $ \sin\theta \gg \max
(1/\gamma,\, 1/\gamma_1)$, the factor $\kappa$ approaches unity.
Then the singlet state in the two--particle c.m.s.
becomes in the laboratory the triplet state
with the zero projection of the
total spin onto the spin rotation axis $z$ and, {\it vice versa}.

Thus, the effect of the relativistic spin rotation leads to the
dependence of the total spin composition
(the singlet and triplet fractions in particular)
on the concrete frame in which the system of two-particles,
moving with different velocity vectors,
is analyzed.
The physical origin of this dependence is the violation
of the parallelism of the spatial axes of the particle rest
frames,
except for the case when the Lorentz transformations to these
frames are done along the directions colinear with the
relative velocity (e.g., from the c.m.s. of the two particles).

We would like to thank M.I. Shirokov for useful remarks.\\
This work was supported by GA Czech Republic, Grant. No. 292/01/0779,
by Russian Foundation for Basic Research, Grant No. 03-02-16210,
and within the Agreements IN2P3-ASCR No. 00-16 and IN2P3-Dubna No.
00-46.


\begin{thebibliography}{99}


\bibitem{lp97}
V.L. Lyuboshitz, M.I Podgoretsky, Yad. Fiz. {\bf 60}, 45
(1997) [Phys. At. Nucl. {\bf 60}, 39 (1997)].
\bibitem{al95}
G.~Alexander, H.J.~Lipkin, Phys. Lett. {\bf B352}, 162 (1995).
\bibitem{led99}
R.~Lednicky, Report MPI-PhE/99-10, Munich (1999).
\bibitem{ll01}
R.~Lednicky, V.L.~Lyuboshitz,
Phys. Lett. {\bf B508}, 146 (2001).
\bibitem{lll03}
R. Lednicky, V.V. Lyuboshitz, V.L. Lyuboshitz,
Yad. Fiz. {\bf 66}, 1007 (2003) [Phys. At. Nucl. {\bf 66},
975 (2003)].

\bibitem{fol56}
L.L. Foldy, Phys. Rev. {\bf 102}, 568 (1956).
\bibitem{sta56}
H.P. Stapp, Phys. Rev. {\bf 103}, 425 (1956).
\bibitem{cs58}
Chou Kuang-chao, M.I. Shirokov, Zh. Eksp.
 Teor. Fiz. {\bf 34}, 1230 (1958)
[Sov. Phys. JETP {\bf 7}, 851 (1958)].
\bibitem{rit61}
V.I. Ritus, Zh. Eksp. Teor. Fiz. {\bf 40}, 352 (1961)
[Sov. Phys. JETP {\bf 13}, 240 (1961)].
\bibitem{shi99}
M.I. Shirokov, Report JINR E2-99-299, Dubna (1999).

\bibitem{bgmr68}
A.M. Baldin, V.I. Gol'dansky, V.M. Maksimenko,
I.L. Rozental',
Kinematika Yadernykh Reakcij (Kinematics of Nuclear
Reactions), Atomizdat, Moscow, 1968, part I, \S 6.

\bibitem{jay78}
B. Jayet et al., Nuovo Cimento {\bf 45A}, 371 (1978).
\bibitem{bar91}
P.D. Barnes et al., Nucl. Phys. {\bf A526}, 575 (1991).
\end{thebibliography}
 \end{document}